\documentclass[12pt]{article}

\usepackage[a4paper,total={18cm,27cm}]{geometry}

\usepackage[utf8]{inputenc}
\usepackage{amsmath}
\usepackage{amsfonts}

\usepackage{graphicx}
\usepackage{array}
\usepackage{caption}
\usepackage{xcolor}
\colorlet{shadecolor}{yellow!20}
\definecolor{symboldarkred}{RGB}{145,0,0}
\newcommand{\redsym}[1]{{\color{symboldarkred}#1}}

\usepackage{authblk}

\usepackage{hyperref}

\usepackage{comment}   

\newcommand{\NI}{\vspace{0.2cm}\noindent}

\newcommand{\is}{\!=\!}

\begin{document}


\title{Surviving by Serving: Functional Relevance Drives Self-Organization in Complex Adaptive Systems}


\author[1]{Claus Metzner}
\author[1]{Ali Ghebleh}
\author[2,3,4]{Achim Schilling}
\author[1]{Andreas Maier}
\author[2,4]{Thomas Kinfe}
\author[1,2,3,4]{Patrick Krauss}

\affil[1]{\small Pattern Recognition Lab, Friedrich-Alexander-University Erlangen-Nürnberg (FAU), Germany}
\affil[2]{\small Neuromodulation and Neuroprosthetics, University Hospital Mannheim, University Heidelberg, Germany}
\affil[3]{\small Neuroscience Lab, University Hospital Erlangen, Germany}
\affil[4]{\small BG Clinic Ludwigshafen, Germany}

\maketitle


\begin{abstract}
\large
\NI Complex adaptive systems often develop organized structures without centralized control. Yet the local mechanisms by which functional organization emerges and persists remain incompletely understood.

Here we propose Surviving by Serving (SBS) as a general principle of self-organization: components persist as long as their outputs are utilized by other components, whereas prolonged non-utilization promotes adaptation and exploration. To investigate this idea, we introduce a minimal multi-agent model in which agents transform shared resources and receive only local feedback when their outputs are subsequently utilized elsewhere in the system.

Despite the absence of global objectives, the system spontaneously self-organizes into functional interaction networks. We observe the emergence of stable transformation chains, core-periphery organization, and the generation of novel states that enable previously inaccessible target conditions to be reached. Remarkably, self-sustaining interaction networks can arise even without external selection pressures, creating a pre-adaptive search phase from which later functional solutions emerge.

These findings suggest that functional utilization may provide a simple, substrate-independent mechanism for the emergence and stabilization of organized structure in complex adaptive systems.

\end{abstract}

\vspace{0.9 cm}

\newpage

\section{Introduction}

\NI Complex adaptive systems often exhibit highly organized structures despite the absence of centralized control \cite{anderson1972more, holland1992complex, mitchell2009complexity}. Examples range from cellular and neural systems to ecosystems, social organizations, and technological networks \cite{barabasi2013network}. In all of these cases, large numbers of interacting components collectively generate coherent global behavior, while each individual component has access only to local information and local interactions.

\NI A central challenge in the study of self-organization is therefore to understand how functional interdependencies emerge and stabilize \cite{camazine2020self, holland1992complex, mitchell2009complexity}. Why do some components become integrated into persistent interaction networks while others disappear, adapt, or remain functionally irrelevant? More generally, which local principles allow organized structures to arise without global planning or supervision? While many existing approaches emphasize mechanisms such as selection, optimization, reinforcement, or autocatalysis \cite{eigen2012hypercycle, kauffman1992origins, hordijk2018autocatalytic}, the local conditions under which functional relevance emerges remain incompletely understood.

\NI In this paper, we argue that ``surviving by serving'' (SBS) may represent a general principle of self-organization. Biological systems provide particularly compelling examples of this idea. They can be understood as dynamical and adaptive networks of many interlocking subsystems \cite{maturana2012autopoiesis, moreno2015biological}. Each component requires certain local conditions in order to function properly, and each component performs specific functions that are essential for other components. Again, it would be virtually impossible for a central control agency to micro-manage all these local activities. Yet such central control is not required if each component has private means to assess and improve the quality of its present local conditions, while also adapting its own output in such a way that other components actually make use of it. Such patterns of mutual dependence and functional closure have been proposed as fundamental organizing principles of living systems \cite{montevil2015biological}.

\begin{figure}[p!]
\centering
\includegraphics[width=0.98\linewidth]{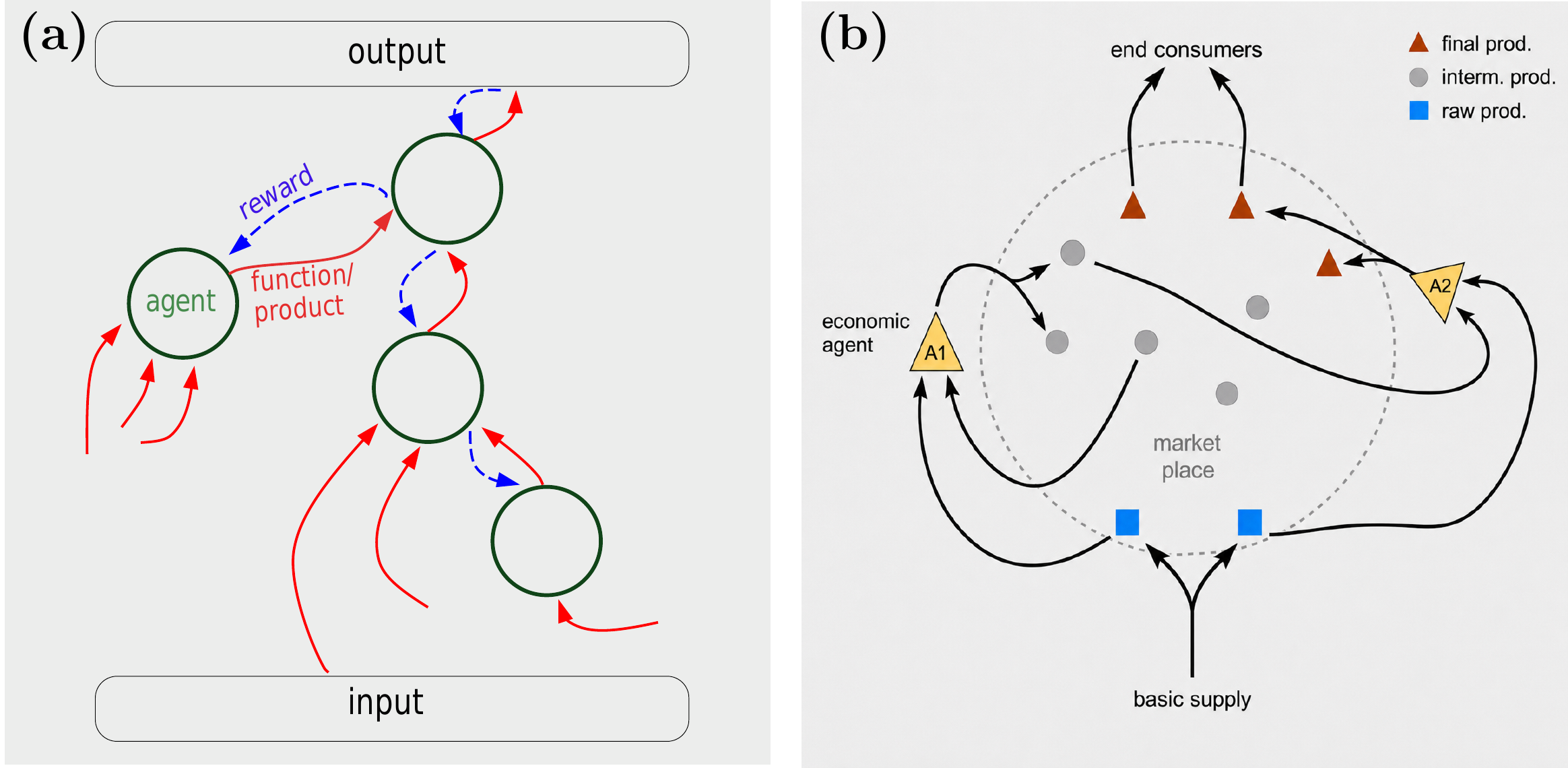}
\caption{
{\bf Main idea and adaptive interaction network.}
{\bf(a)} Principle of ``Surviving by Serving'': Components or agents (green) receive local reinforcement signals (blue) when their outputs, functions, or states (red) are utilized by other components or contribute to externally relevant outcomes. Without continued reinforcement, components remain subject to adaptation.
{\bf(b)} In the model, raw materials (blue squares) are supplied to a shared interaction space (dotted circle), agents (orange triangles) transform suitable input states into new states (gray circles), and evaluators selectively remove states that satisfy predefined target conditions (red triangles).
}
\label{Fig_1}
\end{figure}

\NI This optimization of outputs requires some local form of feedback, reinforcement, or validation from downstream components. In economic systems, money plays the role of such a feedback signal. In biological systems, however, analogous signals may take very different forms. More generally, components within adaptive networks may modify their behavior in ways that increase their usefulness to other components. For example, a neuron whose activity provides functionally relevant information to downstream neurons may become stabilized through synaptic plasticity mechanisms. In this sense, the continued utilization of an output by other components becomes a local indicator that the output has acquired functional value within the larger self-organizing network (Fig.\ref{Fig_1}(a)).

\NI The central idea behind SBS is that persistence is linked to utilization. Components whose outputs repeatedly contribute to ongoing system dynamics acquire functional relevance and become stabilized, whereas components that fail to establish such relations remain subject to continued adaptation. In this way, local interactions can generate a form of distributed selection that operates without centralized control and without explicit knowledge of global system objectives.

\NI To investigate whether this principle is sufficient to generate organized collective behavior, we study a minimal multi-agent system in which components interact exclusively through local transformations and local feedback. The goal is not to model any specific biological, social, or economic system in detail, but rather to examine whether SBS can serve as a general mechanism for the emergence of functional organization in complex adaptive systems.

\NI Specifically, we implement SBS in a highly simplified adaptive interaction network (Fig.\ref{Fig_1}(b)). Agents transform shared resources into new states within a common interaction space and receive feedback only through the subsequent utilization of their outputs by other agents or by external evaluators. This minimal framework allows us to investigate whether utilization-dependent persistence alone is sufficient to generate organized collective behavior, functional interaction networks, and stable patterns of self-organization.

\section{Methods}

\subsection{Simulation overview}

\NI A full simulation run consists of $N_{stp}$ discrete episodes. Each episode, indexed by a discrete time step $t$, internally consists of micro-time steps during which four active system components interact with a shared interaction space in a largely asynchronous manner. These active components are the raw-material source, the aging mechanism, the transformation agents, and the evaluators.

\paragraph{Raw-material source:} At the beginning of each episode, the raw-material source inserts two new raw-material items into the interaction space. Each item belongs to a different state type. These two raw-material types are fixed during initialization and remain unchanged throughout the simulation.

\paragraph{Aging mechanism:} The aging mechanism iterates over all states currently present in the interaction space and removes each of them with a constant probability $P_{dis}$ during every episode.

\paragraph{Transformation agents:} The system contains a temporally fixed number of $N_{agt}$ agents. During each episode, every agent receives one opportunity to perform a transformation step, provided that it finds two input states in the interaction space that match its required properties sufficiently well. Each agent may also adapt its properties when a lack of reinforcement signals indicates that its generated outputs are rarely utilized. The agents are updated in random order.

\paragraph{Evaluators:} The system contains two evaluators. During each episode, each of them has one opportunity to select a suitable state from the interaction space. This can happen only if the state matches the evaluator's target conditions sufficiently well.

\NI Since all four active system components potentially interact with the shared interaction space, its composition generally changes after each episode. The total instantaneous system state can therefore be described by the current composition of the interaction space, the internal properties of the agents, and the cumulative number of states selected by the evaluators.

\subsection{States and their selection}

\NI All raw materials, as well as all required, generated, and target states, are represented by feature vectors in a common continuous space of $N_{dim}$ dimensions. Throughout this paper, we use $N_{dim}=5$. All vectors are normalized to unit length, so that the feature vectors effectively lie on a hypersphere.

\NI To determine whether a given state $\vec{X}$ sufficiently matches the required properties $\vec{Q}_i$ of a transformation agent $i$ or the target conditions $\vec{Q}_k$ of an evaluator $k$, we compute the cosine similarity
\[
S(\vec{X},\vec{Q})=\vec{X} \cdot \vec{Q}.
\]

A state is considered suitable for a given requirement if $S(\vec{X},\vec{Q})$ is greater than or equal to a fixed threshold $\Theta_{thr}$. This threshold controls the selectivity of the adaptive network: low values make many states mutually interchangeable, whereas high values make compatible transformation paths rare.

\subsection{State transformations in agents}

\NI When agent $i$ attempts to perform a transformation step, it first iterates over the states available in the interaction space and constructs the set of candidate states that pass the cosine-similarity threshold with respect to its first required property vector $\vec{Q}^{(A)}_i$. If more than one compatible candidate is available, the first input state is chosen randomly from these candidates. The second input is then chosen analogously with respect to $\vec{Q}^{(B)}_i$, but only among states that also satisfy the exclusion rules described below. Thus input selection is sequential and stochastic; no backtracking is performed if the first random choice prevents a valid second choice. The selected input-state vectors are denoted by $\vec{X}^{(A)}_i$ and $\vec{X}^{(B)}_i$.

\NI If suitable input states are available, agent $i$ removes them from the interaction space, and the generating agents of these states receive one credit point each. Agent $i$ then transforms the inputs into two identical instances of output state $\vec{Y}_i$.

\NI The properties of the output state are computed as follows:
\begin{equation}
\vec{Y}_i =
\operatorname{norm}
\left[
\operatorname{norm}
\left(
a_i \vec{X}^{(A)}_i + b_i \vec{X}^{(B)}_i
\right)
+
\eta_{oos} \operatorname{norm}(\vec{R}_i)
\right].
\label{ProdFormula}
\end{equation}

\NI The two input vectors are first linearly combined (``assembled'') using the agent-specific coefficients $a_i$ and $b_i$ and then normalized. As a result, the output vector $\vec{Y}_i$ depends on the detailed properties of $\vec{X}^{(A)}_i$ and $\vec{X}^{(B)}_i$, although these properties can vary only within the limits imposed by the cosine-similarity filter. However, such a linear combination cannot transcend the subspace spanned by the two inputs.

\NI To allow for this, an agent-specific modification vector $\vec{R}_i$, initialized and mutated randomly, is added to the linear combination, controlled by an out-of-subspace parameter $\eta_{oos}$. When this parameter is much smaller than one, the agent can generate output states with properties that lie slightly outside the subspace of the inputs. This makes longer transformation chains meaningful, because downstream states can carry accumulated transformations from earlier states.

\NI After each transformation step, two copies of the newly generated output state are added to the interaction space, effectively replacing the input states.

\subsection{Input-state exclusion rules}

\NI The two consumed input states of an agent must satisfy exclusion rules that prevent trivial self-use: An agent cannot consume its own output state, the two inputs cannot be the same state instance in the interaction space, and the two inputs cannot originate from the same generating agent. If both inputs are raw materials, they must be of different raw-material types. These constraints force state transformations to depend on relations between different components of the system.

\subsection{Utilization-controlled adaptation of agents}

\NI Agents receive credit when their outputs are utilized. Credit is assigned both when an agent-generated state is consumed as an input by another agent and when it is selected by an evaluator. Credit is not modeled as money. Rather, it is a local reinforcement signal indicating that the agent's current transformation rule has acquired a functional role somewhere in the interaction network.

\NI If an agent repeatedly receives no credit, this indicates that no functional role has been found for the states generated by its current transformation rule. We interpret the resulting change as adaptation under utilization pressure: an unsuccessful agent locally modifies its transformation rule. The probability of adaptation increases with the number $\Delta t$ of consecutive uncredited episodes according to a Hill-type function,
\begin{equation}
P_{ada}(\Delta t)=
\frac{\Delta t^{n}}{\Delta t^{n}+T_{ada}^{n}},
\label{AdaptProb}
\end{equation}
where $T_{ada}$ is the characteristic time scale and $n$ controls the steepness of the transition (Fig.\ref{Fig_2}(a)). This rule tolerates short periods without credit while making persistent lack of utilization increasingly likely to trigger adaptation.

\NI Adaptation slightly mutates the transformation parameters rather than replacing the whole agent. Specifically, when an adaptation event occurs, the two linear combination coefficients $a_i,b_i$ and the components of the modification vector $\vec{R}_i$ are updated by adding independent zero-mean normally distributed random numbers, all with the same standard deviation $\sigma_{ada}$:
\begin{equation}
\begin{aligned}
a_i &\leftarrow a_i+N(0,\sigma_{ada}),\\
b_i &\leftarrow b_i+N(0,\sigma_{ada}),\\
R_{i,k} &\leftarrow R_{i,k}+N(0,\sigma_{ada}).
\end{aligned}
\label{AgentMut}
\end{equation}

\NI The modification vector $\vec{R}_i$ is renormalized after random mutation.

\NI The input-requirement vectors are kept fixed in the current version, although mutation of input preferences would be a natural extension. This local mutation rule preserves partial functionality more gently than complete resampling, while still allowing unsuccessful agents to explore new transformations.

\subsection{State selection by evaluators}

\NI Each of the two evaluators has its own fixed vector $\vec{Q}$ of target conditions. In each episode, it scans the interaction space for agent-generated states that satisfy the cosine-similarity criterion, analogous to the state-selection mechanism of the agents. Raw materials cannot be selected directly by evaluators. If several suitable states are available, one of them is chosen randomly. The evaluator removes the selected state from the interaction space, and the generating agent of that state receives a credit point.

\subsection{State aging}

\NI The aging mechanism iterates over all states currently present in the interaction space, including raw materials, and removes each of them with a fixed probability $P_{dis}$. This prevents the number of states in the interaction space from growing arbitrarily when they are not utilized by any agents or selected by any evaluators.

\subsection{Full-target criterion}

\NI A functional interaction network is said to emerge when both evaluators successfully select suitable states during the same episode. This full-target criterion is deliberately stricter than counting isolated selection events, because it requires the interaction space and transformation network to satisfy all target conditions simultaneously.

\NI For each simulation run, we record the first episode in which the full-target criterion is met, the number of full-target episodes, the total number of selection and transformation events, the size of the interaction space, and the transformation pathways that led to the first successful full-target event.

\subsection{Missing Dimension Hurdle}

\NI In some diagnostic simulations, we introduced a deliberately restrictive initial condition that we refer to as the Missing Dimension Hurdle (MDH). In this condition, both raw-material vectors are constrained to have zero amplitude in feature component 0, whereas the two evaluators depend exclusively on this component. Specifically, the evaluator target vectors are set to opposite directions along the same axis, $(-1,0,0,0,0)$ and $(1,0,0,0,0)$. Thus, no raw material initially contains the feature dimension that is relevant for successful evaluation.

\NI This condition tests whether the adaptive network can generate evaluator-relevant states through chains of input-dependent transformations rather than by directly selecting a pre-existing raw-material feature. Under the transformation rule, a pure linear combination of the raw inputs would remain within the subspace lacking component 0. The exploratory transformation term proportional to $\eta_{oos}\vec{R}_i$ can gradually move states out of this subspace, so successful evaluation requires the transformation dynamics to overcome this geometric hurdle.

\subsection{Temporal sequence of one episode}

\NI Each simulation run starts with a one-time initialization step. The subsequent episode update is then repeated until the chosen simulation horizon of $N_{stp}$ time steps is reached. In the list below, step 0 is performed once, whereas steps 1--7 form the repeated episode loop:

\begin{enumerate}
  \setcounter{enumi}{-1}

  \item \textbf{Initialization:} raw-material vectors, evaluator target vectors, and the initial agent population are drawn in the shared feature space. Each agent receives two input-requirement vectors, two scalar transformation coefficients, and an agent-specific transformation vector. The interaction space is initially empty.

  \item \textbf{State aging:} states already present in the interaction space may disappear with probability $P_{dis}$.

  \item \textbf{External input:} one copy of each raw material is added to the interaction space.

  \item \textbf{State transformation:} agents act once, in random order. If an agent finds two suitable input states, these inputs are consumed and two copies of a transformed output state are added.

  \item \textbf{Evaluation:} each evaluator attempts to select one suitable agent-generated state.

  \item \textbf{Credit assignment:} agents whose outputs were utilized by other agents or selected by evaluators receive credit.

  \item \textbf{Utilization evaluation:} agents without credit accumulate one additional episode without successful utilization.

  \item \textbf{Adaptation:} agents with a sufficiently long period without credit may mutate their transformation coefficients and transformation vector.

\end{enumerate}

\subsection{Transfer-matrix visualization and agent ordering}

\NI To visualize the time-dependent interaction network in the larger simulations, we converted the recorded agent-to-agent state transfers into windowed transfer matrices. For each consecutive time window of 50 episodes, we constructed a matrix $M$ whose entry $M_{ij}$ counts how often a state generated by agent $i$ was used as an input by agent $j$ during that window. Rows and columns therefore represent source and target agents, respectively. Evaluator selections and raw-material inputs were not included in these matrices, because the purpose of the plot was to reveal the internal agent-to-agent interaction network.

\NI The visual order of agents was not chosen independently for each time window. Instead, a single order was computed once from the pre-evaluation interval, $t=1,\ldots,500$, and then kept fixed for all 20 panels of the figure. This makes temporal changes visible without confounding them with changes in axis order. For ordering, we first accumulated the total pre-evaluation transfer matrix and converted it into an undirected similarity matrix by summing reciprocal transfer counts,
\[
C_{ij}=M_{ij}+M_{ji}.
\]
The counts were compressed logarithmically and normalized by the largest observed reciprocal count, yielding
\[
S_{ij}=\frac{\log(1+C_{ij})}{\log(1+C_{\max})}.
\]
We then used $D_{ij}=1-S_{ij}$ as a distance matrix for average-linkage hierarchical clustering. The leaf order of this clustering gives agents with strong reciprocal transfer relations neighboring positions on the matrix axes, so densely interacting groups appear as contiguous blocks.

\NI To make the main core-periphery structure easier to see, we additionally defined a core set from the same pre-evaluation transfer matrix. For each agent, we computed total transfer activity as the sum of all incoming and outgoing agent-to-agent transfer counts during the pre-evaluation interval. Agents were sorted by this activity, and the core was defined as the top-$K$ agents in this ranking. The value of $K$ was chosen by visual inspection of the pre-evaluation transfer structure and then fixed for the subsequent time-window and adaptation-rate analyses.

\subsection{Initialization rules}

\NI Initial random feature vectors are generated by drawing each component independently from a uniform distribution on $[-1,1]$ and normalizing the resulting vector. In addition, several initial vector pairs are constrained to be sufficiently dissimilar. Specifically, the two raw-material vectors, the two evaluator-target vectors in unconstrained simulations, and the two input-requirement vectors of each agent are generated as diverse pairs: after the first vector has been drawn, the second vector is redrawn up to 1000 times until its cosine similarity to the first vector is below 0.45. If no candidate satisfies this condition, the candidate with the lowest cosine similarity among the attempted draws is used. The agent-specific modification vector $\vec{R}_i$ is initialized in the same random-vector ensemble but without an additional pairwise-diversity constraint. The initial linear-combination coefficients $a_i$ and $b_i$ are drawn independently from a uniform distribution on $[0,1]$; later Gaussian adaptation steps are not clipped to this interval.

\subsection{Summary of symbols}

\NI The symbols used in the model description are summarized in Table~\ref{Tab_MethodSymbols}.

\begin{table}[ht]
\centering
\renewcommand{\arraystretch}{1.18}
\begin{tabular}{|>{\raggedright\arraybackslash}p{0.25\linewidth}|>{\raggedright\arraybackslash}p{0.67\linewidth}|}
\hline
\textbf{Symbol} & \textbf{Meaning} \\
\hline
$N_{stp}$ & Number of simulation time steps (episodes). \\
\hline
$N_{dim}$ & Dimension of the continuous feature space. \\
\hline
$N_{raw}=2$ & Number of raw-material state types supplied to the interaction space. \\
\hline
$N_{agt}$ & Number of agents in the adaptive network. \\
\hline
$N_{con}=2$ & Number of evaluators. \\
\hline
$\vec{Q}$ & Requirement vector used to define cosine similarity. \\
\hline
$\Theta_{thr}$ & Similarity threshold above which a state is suitable for a requirement. \\
\hline
$\eta_{oos}$ & Out-of-subspace mixing factor controlling the contribution of $\vec{R}_i$ to state transformations. \\
\hline
$P_{dis}$ & Probability that an existing state disappears from the interaction space at the beginning of an episode. \\
\hline
$P_{ada}(\Delta t)$ & Probability that an agent adapts after $\Delta t$ consecutive uncredited episodes. \\
\hline
$T_{ada}=30$ & Characteristic time constant of adaptation. \\
\hline
$n=4$ & Hill exponent controlling the steepness of the adaptation-probability function. \\
\hline
$\sigma_{ada}=0.05$ & Standard deviation of the Gaussian mutation applied during adaptation. \\
\hline
$\redsym{\vec{X}}$ & Generic state vector used to define cosine similarity. \\
\hline
$\redsym{a_i}$, $\redsym{b_i}$ & Scalar transformation coefficients of agent $i$. \\
\hline
$\redsym{\vec{R}_i}$ & Agent-specific transformation vector of agent $i$. \\
\hline
$\redsym{\vec{X}^{(A)}_i}$, $\redsym{\vec{X}^{(B)}_i}$ & Input-state vectors consumed by agent $i$. \\
\hline
$\redsym{\vec{Y}_i}$ & Output-state vector generated by agent $i$. \\
\hline
$\redsym{\Delta t}$ & Number of consecutive episodes in which an agent has received no credit. \\
\hline
\end{tabular}
\caption{Summary of symbols used in the Methods section. Black symbols in the upper part denote fixed system-control parameters. Dark-red symbols denote dynamically changing state variables or agent quantities modified by adaptation.}
\label{Tab_MethodSymbols}
\end{table}

\section{Results}

\subsection{Emergence of a functional interaction network}

\NI As in all simulations presented in this paper, we use a state feature space with $N_{dim}\is5$ dimensions, $N_{raw}\is2$ types of raw materials, and $N_{con}\is2$ evaluators. We first consider a run containing only $N_{agt}\is5$ agents, in order to test whether and when a functional interaction network self-organizes that can generate suitable states for both evaluators within the same episode. We refer to this situation as the full-target criterion.

\NI Transformation agents and evaluators compare the available states with their own requirements or target conditions, respectively, as described in Methods section 2.2. A state is accepted if the cosine similarity of the corresponding feature vectors exceeds the threshold $\theta_{acc}$.

\NI At the beginning of the simulation, only raw materials are available in the interaction space. Consequently, the first successful transformation event, if it occurs, must involve an agent that accepts the two raw-material states as inputs. This is not guaranteed: if all initial agent-input similarities to the raw-material states fall below the acceptance threshold, no agent can transform immediately, and stochastic adaptation is first required before a raw-material-based transformation becomes possible.

\begin{figure}[p!]
\centering
\includegraphics[width=0.80\linewidth]{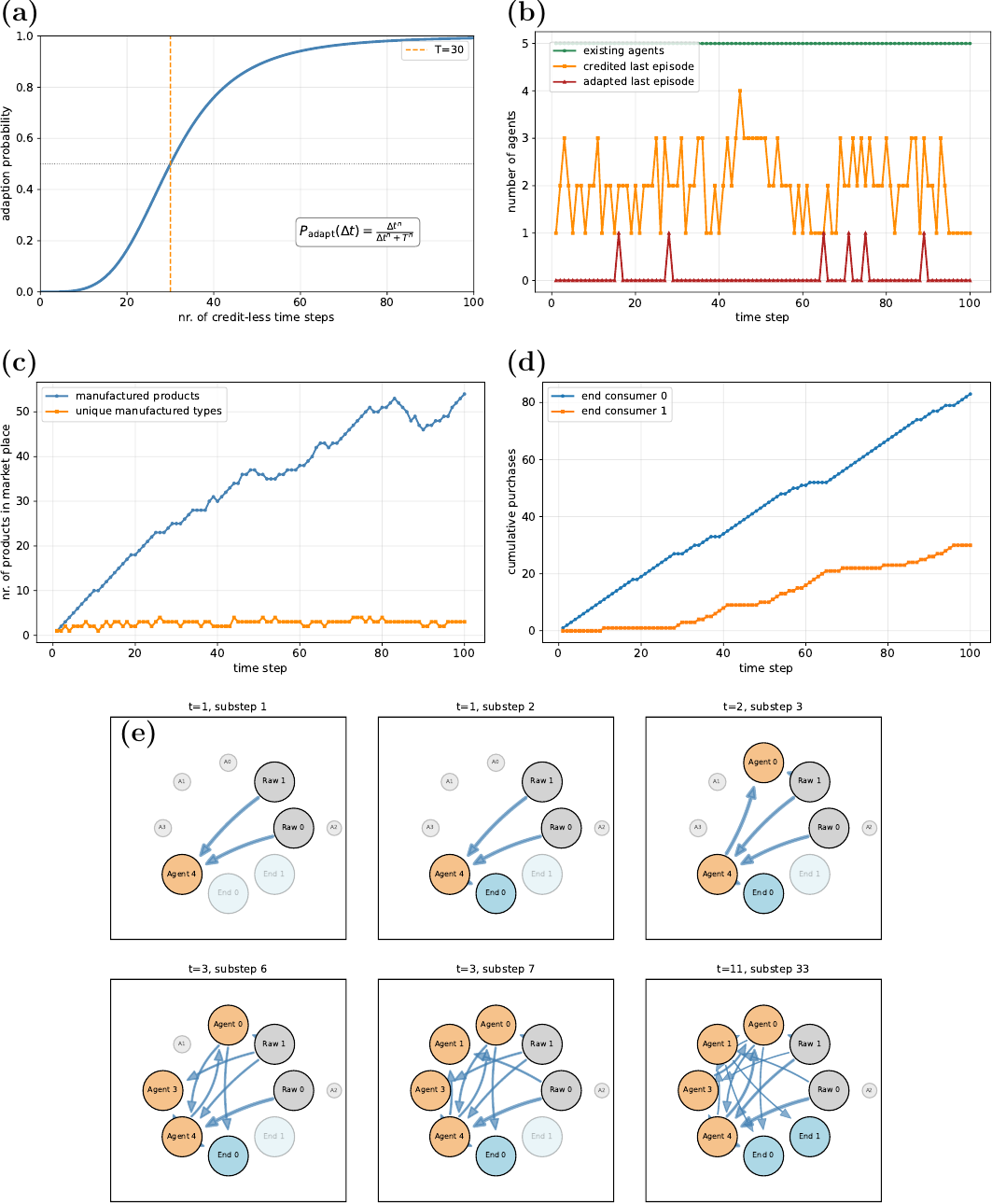}
\caption{
{\bf Emergence of a functional interaction network.}
{\bf(a)} The probability of adaptation increases with the number of consecutive credit-less time steps according to the Hill-type rule used in the simulation.
{\bf(b)} Existing agents, agents credited in the previous episode, and agents adapted in the previous episode over time.
{\bf(c)} Number of generated states and unique state types present in the interaction space.
{\bf(d)} Cumulative state selections by the two evaluators.
{\bf(e)} Stepwise emergence of the first full-target situation, in which both evaluators successfully select suitable states within the same episode. Snapshots were selected at structural transition points: a new panel was added when an agent first obtained two distinct input suppliers or when an evaluator successfully selected a state for the first time. Each panel shows cumulative transfers up to that substep, ending with the first episode in which both evaluators are simultaneously satisfied.
Used parameters: MDH=False, $N_{agt}=5$, $N_{raw}=2$, $N_{con}=2$, $N_{dim}=5$, $N_{stp}=220$, $\Theta_{thr}=0.48$, $\eta_{oos}=0.8$, $P_{dis}=0.01$, $T_{ada}=30$, $n=4$, $\sigma_{ada}=0.05$, evaluator activity enabled, random seed $=12$.
}
\label{Fig_2}
\end{figure}

\NI In the specific run shown in Fig.~\ref{Fig_2}, however, the relatively permissive threshold $\theta_{acc}\is0.48$ allows agent 4 to accept both raw-material states already during the first episode, as shown in the upper-left snapshot of Fig.~\ref{Fig_2}(e). Even more noteworthy, the state generated by agent 4 is already accepted by evaluator 0 in the same first episode.

\NI The full-target criterion, however, is reached only after 10 additional episodes, as shown in the lower-right snapshot of Fig.~\ref{Fig_2}(e). In this final snapshot, evaluator 1 selects a state generated by agent 1. Importantly, agent 1 does not simply combine the two raw materials, but utilizes an upstream state generated through a short transformation chain involving agents 4 and 0\footnote{Note that the snapshots in Fig.~\ref{Fig_2}(e) were chosen according to an event-based criterion rather than by uniform temporal spacing. We followed all state-transfer events up to the first full-target episode and added a snapshot whenever the cumulative interaction graph acquired a new functional milestone: either an agent became connected to two distinct input suppliers for the first time, or a previously unsatisfied evaluator successfully selected a state for the first time. Each panel therefore shows the cumulative transfer network up to the corresponding substep. The sequence thus highlights structural changes in the emerging interaction network rather than equal intervals in simulation time.}.

\NI This first run therefore demonstrates that self-organization of functional interaction networks is indeed possible, based on the local principle of surviving by serving.

\NI On timescales of up to 100 episodes, we observe that the total number of states in the interaction space still tends to increase and, despite the aging mechanism, shows no clear signs of saturation (Fig.~\ref{Fig_2}(c), blue curve). However, many of these states are identical or similar outputs generated by the same few agents, as the number of unique state types remains below five (Fig.~\ref{Fig_2}(c), orange curve). In particular, once the interaction network has formed, it remains stable and enables sustained evaluator selections (Fig.~\ref{Fig_2}(d)).

\NI We find that, of the five available agents, only a core group of one to four agents receives credit in any given episode, meaning that the outputs of the remaining agents are not utilized within the adaptive network (Fig.~\ref{Fig_2}(b), green and orange curves). Continued lack of credit increases the adaptation probability of agents according to a Hill-type function (Fig.~\ref{Fig_2}(a)). Indeed, we observe six cases of agent adaptation during the first 100 episodes (Fig.~\ref{Fig_2}(b), red curve).

\subsection{Overcoming the Missing Dimension Hurdle}

\NI Generating novel states from the inputs currently available in the interaction space can be viewed as a creative process in which qualitatively new properties emerge that were not present in the original inputs. The evaluators may specifically favor such novel properties.

\NI We simulate this generation of novelty by deliberately providing only raw materials whose feature vectors have no 0th component. We refer to this situation as the Missing Dimension Hurdle (MDH). At the same time, we impose target conditions that depend exclusively on this missing feature dimension by setting the evaluator target vectors to $(-1,0,0,0,0)$ and $(+1,0,0,0,0)$, respectively.

\NI In our model system, agents first form a linear combination of the input feature vectors (Eq.~\ref{ProdFormula}). With this operation alone, the output state could never leave the subspace spanned by the input states. However, after the linear combination, each agent adds a small agent-specific feature vector that generally has non-zero entries in all five components. Through the normalization and weighting steps in Eq.~\ref{ProdFormula}, we can control the degree of this out-of-subspace transformation by the parameter $\eta_{oos}$. For small values of $\eta_{oos}$, each agent can generate only tiny traces of novel properties, but the interaction of several agents may amplify these properties until the resulting states satisfy the evaluator target conditions.

\begin{figure}[p!]
\centering
\includegraphics[width=0.88\linewidth]{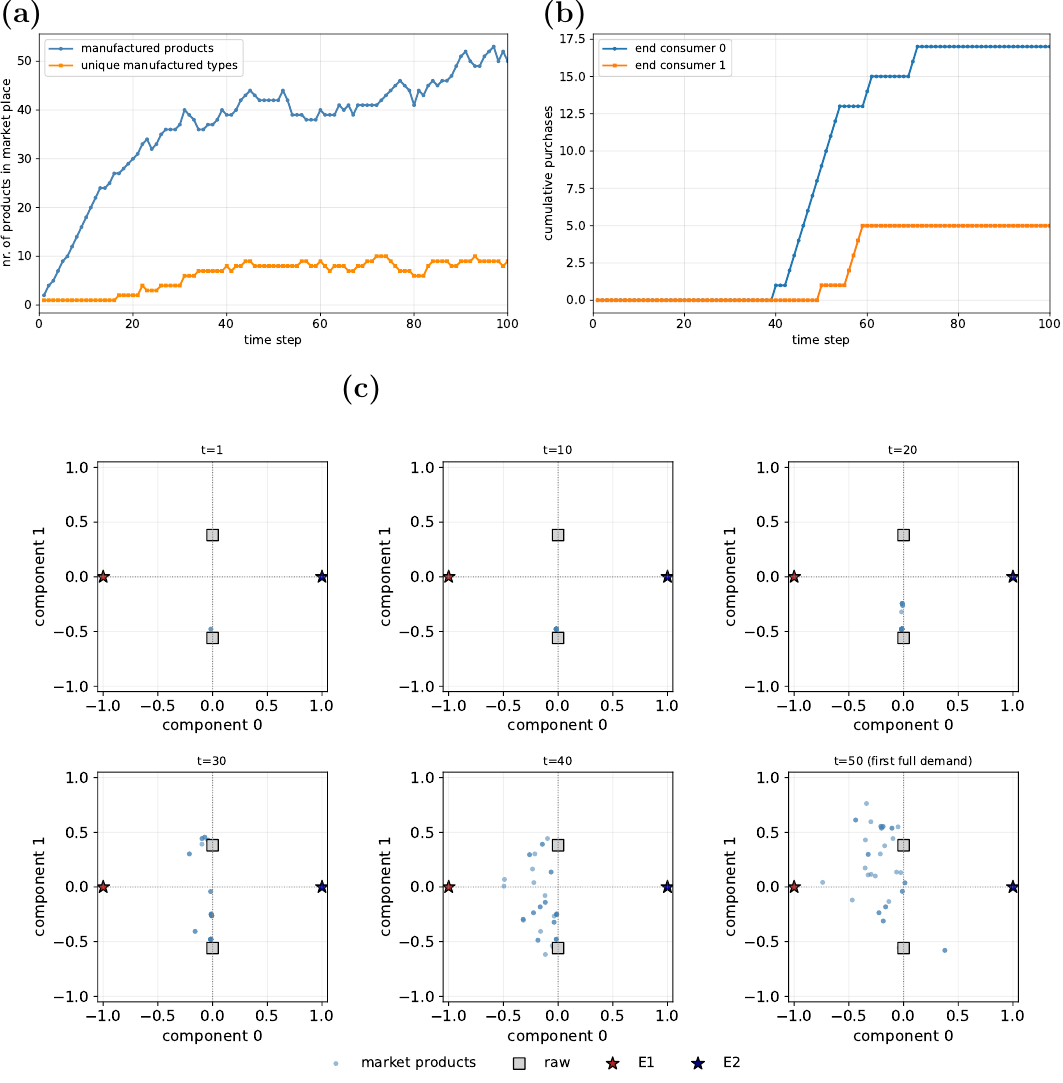}
\caption{
{\bf Overcoming the Missing Dimension Hurdle.}
{\bf(a)} Number of generated states and unique state types present in the interaction space for the component-0-constrained diagnostic run.
{\bf(b)} Cumulative state selections by the two evaluators.
{\bf(c)} State vectors in the interaction space projected onto components 0 and 1 at selected times up to the first full-target episode. Raw materials start with zero component-0 amplitude, whereas the two evaluators impose opposite target conditions along component 0.
Used parameters: MDH=True, $N_{agt}=10$, $N_{raw}=2$, $N_{con}=2$, $N_{dim}=5$, $N_{stp}=500$, $\Theta_{thr}=0.48$, $\eta_{oos}=0.05$, $P_{dis}=0.03$, $T_{ada}=30$, $n=4$, $\sigma_{ada}=0.05$, random seed $=214$.
}
\label{Fig_3}
\end{figure}

\NI To test this mechanism, we use a slightly larger system of $N_{agt}\is10$ agents (Fig.~\ref{Fig_3}). In this case, the simulation generates an interaction space containing about 50 states during the first 100 episodes (Fig.~\ref{Fig_3}(a), blue curve), a number comparable to the previous experiment (Fig.~\ref{Fig_2}(c), blue curve). However, the number of unique states simultaneously present in the interaction space is now about 10, which is considerably larger than before.

\NI Interestingly, it now takes almost 40 episodes before the first evaluator successfully selects a state. This already suggests that many adaptive processes are required before the collective dynamics can generate states with large amplitudes in the 0th component.

\NI To visualize this adaptation process and the gradual exploration of state space, we consider only the components $C_0$ and $C_1$ of the feature vectors. For selected episodes $t$, each state currently present in the interaction space, including the two raw materials, is represented as a dot within the $(C_0,C_1)$ feature plane (Fig.~\ref{Fig_3}(c)).

\NI During the first 30 episodes, all states are still strongly concentrated around the axis $C_0\is0$, indicating that the novel property is present only in tiny traces. Subsequently, however, the point cloud rapidly diffuses into regions with non-zero $C_0$, until the first full-target condition is reached at $t\is50$.

\subsection{Formation of a core network}

\NI In the experiments discussed above, we found that only a fraction of the agents receive credit in each episode, while many similar states accumulate in the interaction space. Taken together, these observations suggest that a small core network may dominate the interaction dynamics. Since the members of this group receive regular credit, they have little incentive to adapt, which can lead to an effective self-stabilization of the configuration. By contrast, agents outside this core network, referred to below as peripheral agents, remain largely uncredited and therefore continue to adapt.

\NI If true, this raises several questions. To what extent does the core group always consist of the same agents? Is the core a single highly interconnected cluster, or is it structured into multiple sub-cores? How long does the core remain stable? And how is it affected by external evaluation?

\NI We address these questions with a slightly larger network of $N_{agt}\is24$ agents and a longer simulation time of $N_{stp}\is1000$ episodes. The simulation is performed under the Missing Dimension Hurdle condition. During the first half of the simulation, however, evaluator activity is switched off.

\NI The pairwise coupling strength between two agents $i$ and $j$ is quantified by the number of state transfers from $i$ to $j$ within a given time window. Such a transfer also implies credit assignment in the reverse direction, from $j$ to $i$. The internal interaction dynamics of the agent collective, excluding raw-material inputs and evaluator selections, can therefore be represented by a transfer matrix for each time window.

\begin{figure}[p!]
\centering
\includegraphics[width=0.80\linewidth]{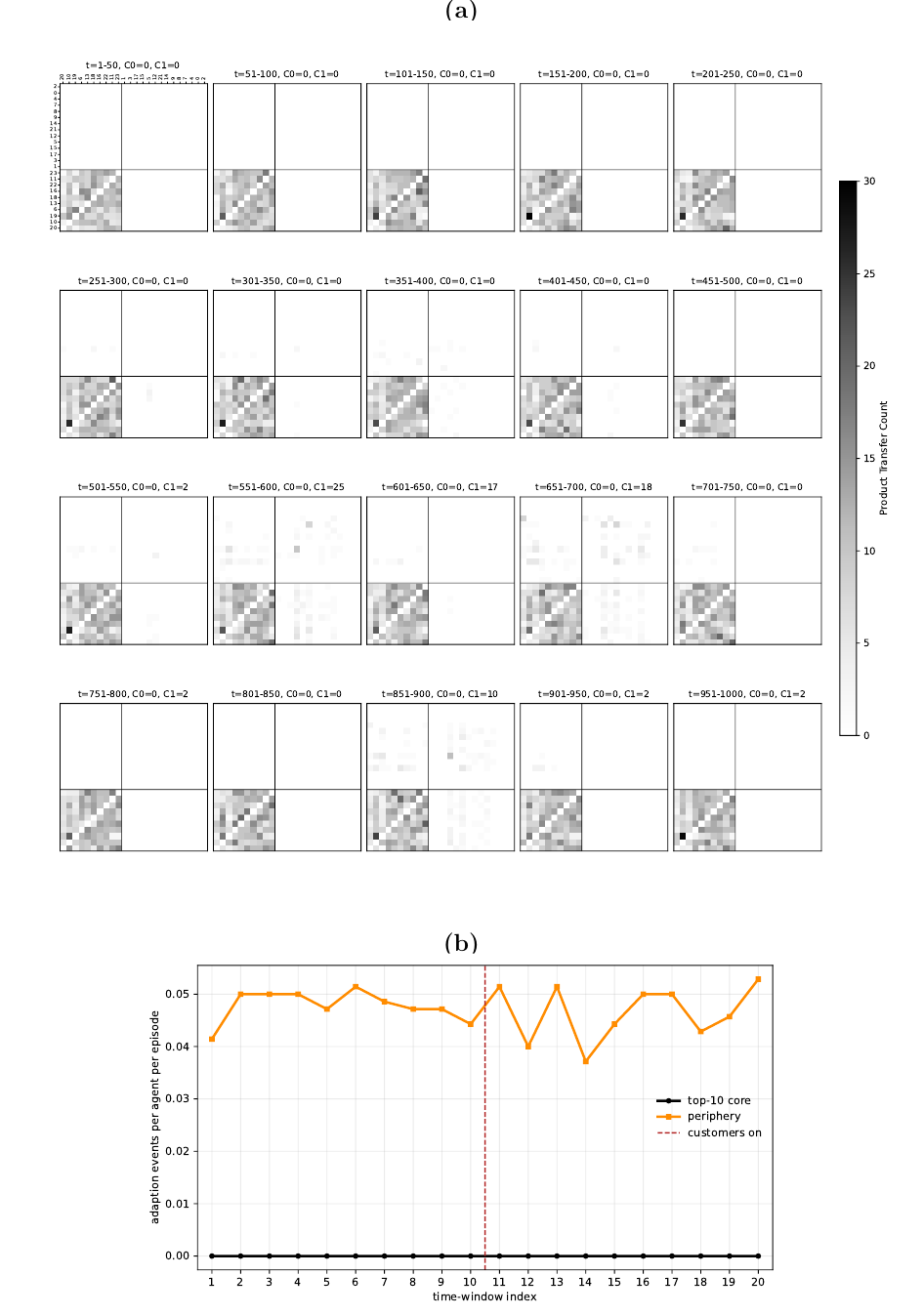}
\caption{
{\bf Formation of a Core Network (Strong Cluster Example).}
{\bf(a)} Counts of pairwise state transfers between agents, represented by gray values, are shown in consecutive 50-episode windows. In the simulation, evaluator activity is disabled during the first 500 episodes and enabled from episode 501 onward.
The title of each matrix shows the time window and the numbers of selections by the two evaluators $C0$ and $C1$ during that interval.
Agent order is determined by hierarchical clustering based on the complete pre-evaluation interval $t=1,\ldots,500$ and fixed for all matrices.
The black separator marks the top-$K$ core boundary with $K=10$ agents, selected according to their transfer activity during the pre-evaluation interval.
{\bf(b)} Adaptation rate in the top-10 core and in the periphery, computed in the same 50-episode windows and normalized by group size and window length.
Used parameters: MDH=True, $N_{agt}=24$, $N_{raw}=2$, $N_{con}=2$, $N_{dim}=5$, $N_{stp}=1000$, $\Theta_{thr}=0.20$, $\eta_{oos}=0.05$, $P_{dis}=0.01$, $T_{ada}=30$, $n=4$, $\sigma_{ada}=0.05$, random seed $=8$.
}
\label{Fig_4}
\end{figure}

\NI In a first run, we find that a well-defined core of 10 active agents emerges already within the first 50-episode time window (Fig.~\ref{Fig_4}(a), top-left matrix). There is no apparent structure within this core that would indicate pronounced sub-clusters of activity. Outside the core, we observe only very little transfer activity among peripheral agents or between the periphery and the core.

\NI Remarkably, the composition of the core remains stable across all 20 subsequent time windows, although the state fluxes within the core fluctuate. Activity in the periphery slightly increases in some time windows, but it never reaches the level observed within the core.

\NI It is equally interesting that enabling evaluator activity after $t\is500$ does not significantly affect the transfer dynamics of the system.

\NI Although the network evolved without external evaluation during the first 10 time windows, evaluator C1 begins selecting states already in the first window after evaluation is enabled. This suggests that the self-organized network has already generated a sufficiently diverse repertoire of states, so that some externally imposed target conditions can be satisfied without substantial further adaptation. In this case, these target conditions depend on large magnitudes in the 0th feature component due to the Missing Dimension Hurdle condition. By contrast, the second evaluator does not perform a single successful selection throughout the entire simulation run.

\begin{figure}[p!]
\centering
\includegraphics[width=0.80\linewidth]{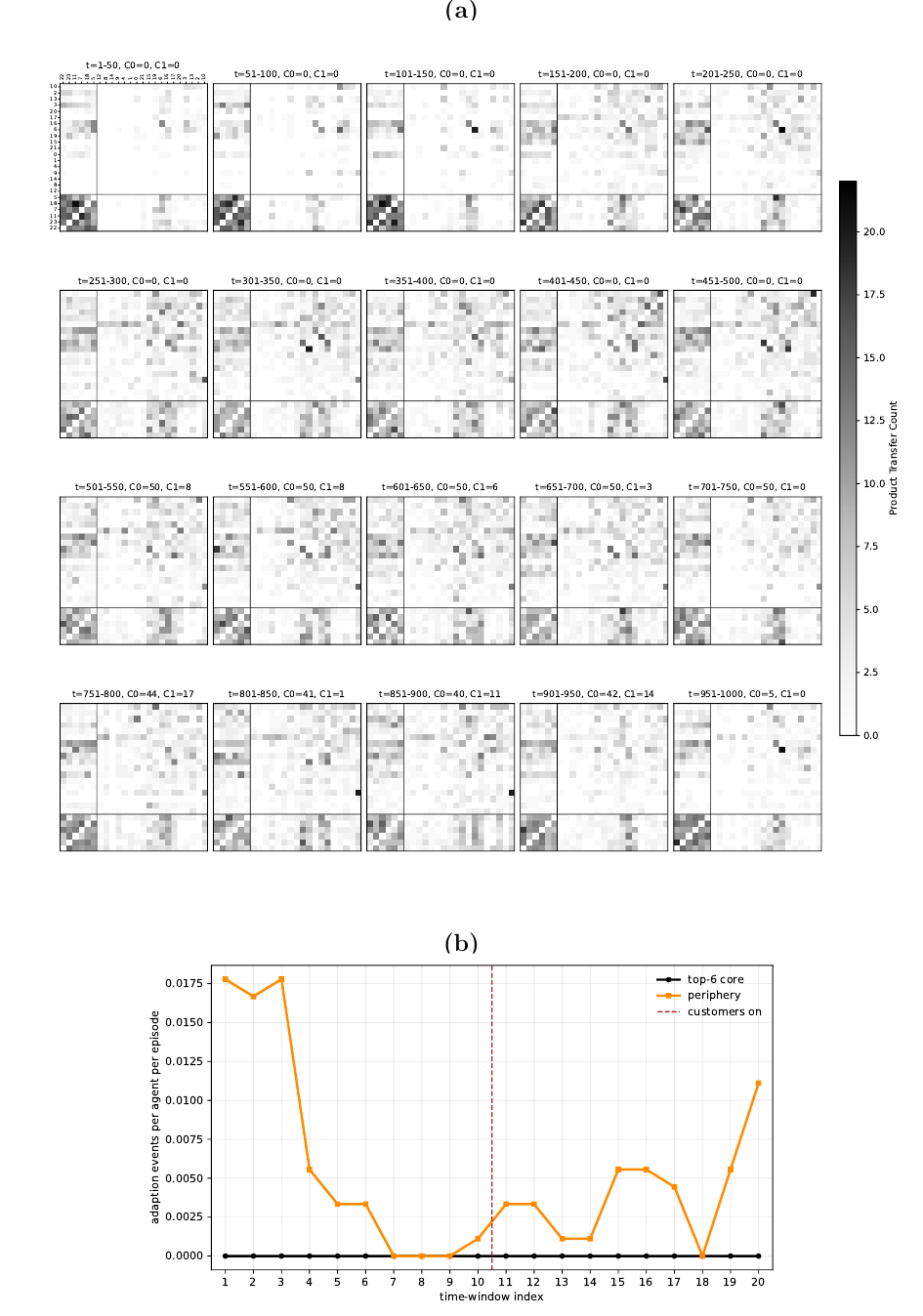}
\caption{
{\bf Formation of a Core Network (Weaker Cluster Example).}
{\bf(a)} Counts of pairwise state transfers between agents, represented by gray values, are shown in consecutive 50-episode windows. In the simulation, evaluator activity is disabled during the first 500 episodes and enabled from episode 501 onward.
The title of each matrix shows the time window and the numbers of selections by the two evaluators $C0$ and $C1$ during that interval.
Agent order is determined by hierarchical clustering based on the complete pre-evaluation interval $t=1,\ldots,500$ and fixed for all matrices.
The black separator marks the top-$K$ core boundary with $K=6$ agents, selected according to their transfer activity during the pre-evaluation interval.
{\bf(b)} Adaptation rate in the top-6 core and in the periphery, computed in the same 50-episode windows and normalized by group size and window length.
Used parameters: MDH=True, $N_{agt}=24$, $N_{raw}=2$, $N_{con}=2$, $N_{dim}=5$, $N_{stp}=1000$, $\Theta_{thr}=0.20$, $\eta_{oos}=0.05$, $P_{dis}=0.01$, $T_{ada}=30$, $n=4$, $\sigma_{ada}=0.05$, random seed $=85$.
}
\label{Fig_5}
\end{figure}

\NI In a second run (Fig.~\ref{Fig_5}(a)), which differs from the first only in the random seed, we observe qualitatively similar behavior. In this case, however, the difference between core and periphery is considerably less pronounced, and the overall transfer structure is temporally much less stable. In this run, both evaluators begin selecting states immediately after evaluation is enabled.

\NI In the two simulation runs discussed above, we counted not only state transfers between agents within each time window $t$, but also adaptation events. After normalizing these counts by the number of agents and by the length of the time window, we plotted them as a function of $t$ in Fig.~\ref{Fig_4}(b) and Fig.~\ref{Fig_5}(b), separately for the core and the periphery.

\NI In both runs, we find that no adaptations take place within the core (black curves), whereas the periphery exhibits an ongoing but fluctuating adaptation rate (orange curves). This supports the original hypothesis that the core network is protected from adaptive changes by regular credited exchanges among its members, while peripheral agents continue to search for functional roles within the adaptive network.

\subsection{Effect of the acceptance threshold on the adaptive network}

\NI In the experiments of the previous section, the adaptive network was able to autonomously generate states that satisfied externally imposed target conditions without prior adaptation toward this specific goal. This is, however, far from guaranteed. In particular, when the acceptance threshold $\Theta_{acc}$ is raised, matching between actual and required state properties becomes increasingly difficult, and the formation of functional interaction networks is hampered.

\begin{figure}[p!]
\centering
\includegraphics[width=0.90\linewidth]{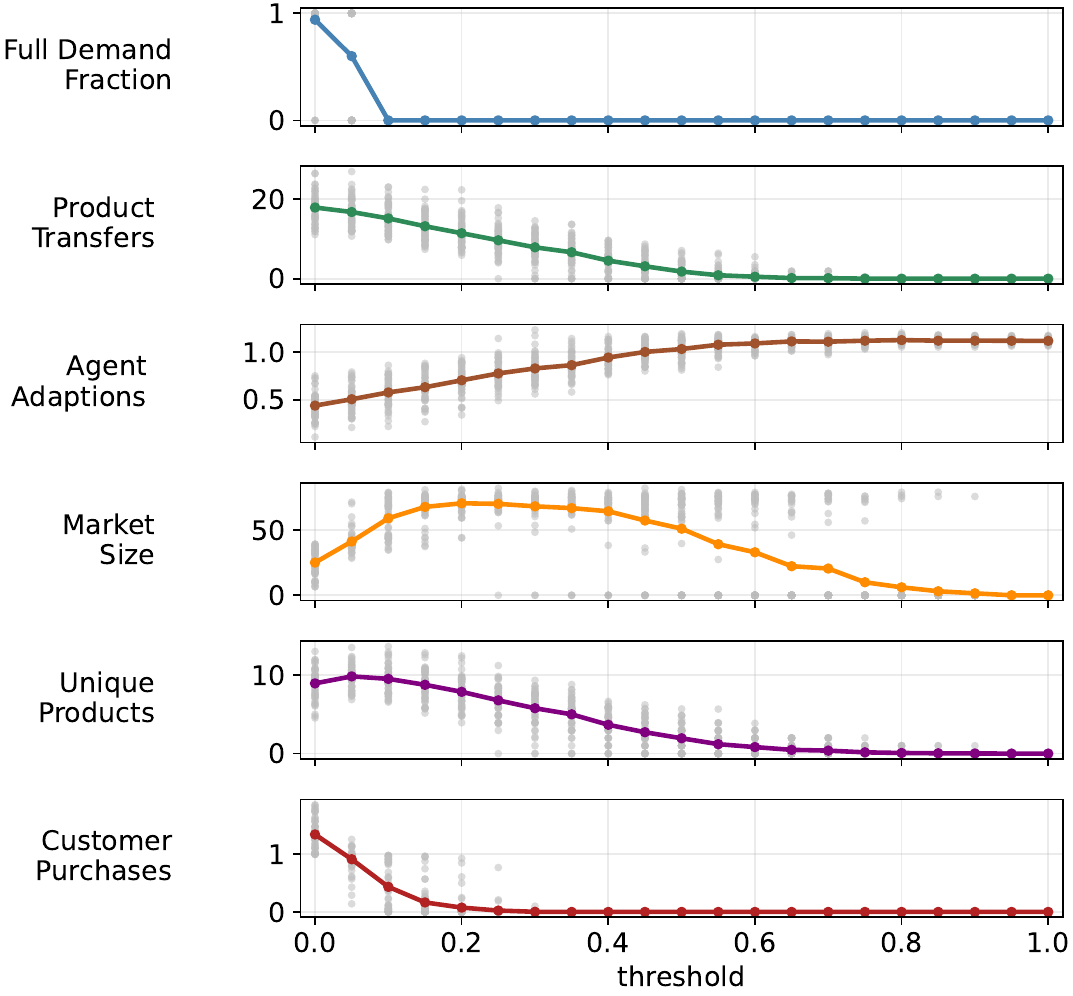}
\caption{
{\bf Effect of the acceptance threshold on the adaptive network.}
Several key indicators of the collective dynamics are computed as a function of the acceptance threshold $\Theta_{acc}$. Results are averaged over 50 repetitions for each value of $\Theta_{acc}$, using the same 50 random seeds each time. Evaluator activity was enabled from the beginning of each run.
From top to bottom, the panels show the fraction of runs reaching the full-target criterion, the mean number of agent-to-agent state transfers per episode, the mean number of agent adaptations per episode, the mean number of generated states present in the interaction space, the mean number of unique state types, and the mean number of evaluator selections per episode.
Gray dots denote individual repeated runs, whereas colored curves show averages over repetitions for each threshold value.
Used parameters: MDH=True, $N_{agt}=24$, $N_{raw}=2$, $N_{con}=2$, $N_{dim}=5$, $N_{stp}=100$, $N_{rep}=50$, $\eta_{oos}=0.05$, $P_{dis}=0.01$, $T_{ada}=30$, $n=4$, $\sigma_{ada}=0.05$.
}
\label{Fig_6}
\end{figure}

\NI To test the influence of $\Theta_{acc}$ on the adaptive network, we use the same control parameters as in the previous experiment, except that evaluator activity is enabled from the beginning and that each run lasts only 100 episodes. The acceptance threshold $\Theta_{acc}$ is swept from 0 to 1, and 50 independent runs are performed for each chosen value of $\Theta_{acc}$ (Fig.~\ref{Fig_6}).

\NI We find that the fraction of runs in which both evaluators successfully select states during the same episode at least once, corresponding to fulfillment of the full-target criterion, decreases very rapidly from 1 to 0 with increasing $\Theta_{acc}$, as expected (top panel, blue curve). A qualitatively similar behavior is observed for the mean number of evaluator selections per episode (bottom panel, red curve).

\NI As $\Theta_{acc}$ is raised, matching becomes more difficult. Consequently, the mean number of agent-to-agent state transfers per episode gradually decreases (second panel from the top, green curve), while the mean number of agent adaptations per episode increases (third panel from the top, brown curve), since agents cannot earn credit without successful state transfers.

\NI Interestingly, both the size of the interaction space, measured as the mean number of generated states present in the system (fourth panel from the top, orange curve), and state diversity, measured as the mean number of states originating from different agents (fifth panel from the top, magenta curve), show a peak-like dependence on $\Theta_{acc}$. However, the peak positions are not identical for the two quantities.

\section{Discussion}

\subsection{Summary}

\NI The central motivation of this work was to test whether a simple local principle, which we call surviving by serving (SBS), can generate organized collective behavior without centralized control. For this purpose, we introduced a minimal adaptive network in which raw materials enter a shared interaction space, agents transform suitable pairs of states into new states, and evaluators selectively remove states that satisfy predefined target conditions. Agents do not possess any knowledge of global system objectives. Instead, they receive only local credit (local reinforcement signal) when their outputs are utilized by other agents or selected by evaluators, while prolonged absence of reinforcement increases their probability of adaptation.

\NI We found that this local SBS rule can indeed give rise to the self-organized emergence of functional interaction networks. Even in small systems, agents form transformation chains that progressively generate states capable of satisfying external target conditions. Thus, organized collective behavior can emerge although each agent reacts only to local utilization or non-utilization of its own outputs.

\NI The Missing Dimension Hurdle experiment further demonstrated the system's capacity to generate novelty. When the raw materials initially lack the feature dimension required for successful evaluation, the system must leave the original raw-material subspace in order to satisfy the target conditions. Through repeated transformations and adaptive exploration, states gradually diffuse through feature space until previously inaccessible regions become reachable and novel functionally relevant states emerge.

\NI A particularly unexpected result was the spontaneous formation of a stable core surrounded by a more weakly active adaptive periphery. Agents within the core exchange states with one another and thereby provide sufficient mutual reinforcement to suppress further adaptation. This creates a self-stabilizing interaction network. Remarkably, such cores can emerge even before external evaluation becomes active, and their internal dynamics may remain largely independent of subsequent evaluator activity. This suggests that internally circulating interaction networks can act as a preadaptive search phase that maintains activity, explores state space, and may later contain states or transformation chains capable of satisfying externally imposed target conditions.

\NI Finally, parameter sweeps revealed that the collective dynamics depend systematically on the acceptance threshold. Increasing this threshold reduces compatibility between states, decreases transfer activity and successful evaluations, and increases adaptation pressure. At the same time, both the size of the interaction space and state diversity exhibit non-monotonic dependencies, indicating that intermediate selectivity may promote particularly rich internal repertoires of states and interactions.

\subsection{SBS as a General Principle of Self-Organization}

\NI The central contribution of the present work is not the specific agent-based model itself, but the more general hypothesis that organized structure can emerge from a simple local principle: components persist when their outputs are utilized by other components. We refer to this principle as surviving by serving (SBS). In its most abstract form, SBS links persistence to functional relevance. Components whose outputs repeatedly contribute to ongoing system dynamics become stabilized, whereas components that fail to establish such functional relations remain subject to continued adaptation.

\NI Importantly, SBS does not require centralized control, global optimization, or explicit knowledge of system-wide objectives. Each component reacts only to local information, namely whether its outputs continue to be utilized within the surrounding network. Nevertheless, our simulations demonstrate that such purely local interactions can give rise to highly organized collective behavior, including stable interaction networks, core-periphery structures, adaptive exploration, and the emergence of novel states.

\NI From this perspective, SBS can be interpreted as a distributed mechanism of functional selection. Rather than selecting components based on predefined fitness functions or externally specified goals, the system continuously evaluates components through their actual participation in ongoing interactions. Components that contribute to the activity of other components remain stable, while components that fail to do so continue to explore alternative configurations. In this way, functional organization emerges as a consequence of network dynamics rather than external design.

\NI The resulting picture differs from many traditional accounts of self-organization. Classical approaches often emphasize optimization, equilibrium formation, energy minimization, or adaptation to external selection pressures. By contrast, SBS focuses on the relational role of components within a network. What matters is not the intrinsic nature of a component, but whether its outputs become integrated into recurrent patterns of interaction. Functional relevance therefore emerges from utilization itself.

\NI The simulations presented here suggest that this simple principle may already be sufficient to generate several hallmark phenomena of complex adaptive systems. Stable structures emerge without centralized planning, exploratory dynamics coexist with robust functional organization, and internally sustained interaction networks can form even before external evaluation becomes active. Taken together, these observations support the view that SBS may represent a general mechanism through which complex systems self-organize into functionally coherent structures.

\subsection{Autocatalysis, Autopoiesis and Organizational Closure}

\NI A particularly close conceptual connection exists between SBS and theories of self-sustaining organization. Hypercycle theory and Kauffman's work on autocatalytic networks demonstrated how mutually supporting reaction systems can acquire collective persistence without external design \cite{eigen1979hypercycle,kauffman1993origins}. More recently, the formal theory of reflexively autocatalytic and food-generated (RAF) sets showed how networks of reactions can become collectively self-maintaining when their components recursively support one another and are ultimately grounded in an external food set \cite{hordijk2004detecting,hordijk2012autocatalytic,hordijk2018autocatalytic}.

\NI The present model shares important similarities with these approaches. Raw materials enter the system from an external source, while agents transform existing states into new states that may subsequently become useful to other agents. As a result, self-sustaining networks of mutually supporting transformations can emerge. However, SBS introduces an additional dynamical principle. Persistence is not determined solely by catalytic closure or network topology, but by continued utilization. Components that are repeatedly used become stabilized, whereas components that fail to establish functional relations remain subject to adaptation and exploration.

\NI SBS is also closely related to theories of biological autonomy and organizational closure. Autopoietic theory emphasized that living systems maintain themselves through networks of processes that continuously regenerate the conditions of their own existence \cite{maturana1980autopoiesis,varela1979principles}. More recent work on biological autonomy and the closure of constraints has described organisms as systems in which mutually dependent constraints stabilize one another across time \cite{moreno2015biological,montevil2015biological}.

\NI Although the present model is not autopoietic in the strict biological sense, it exhibits a related organizational feature. Agents that generate states useful to other agents receive continued reinforcement and therefore remain stable. Their outputs contribute to the persistence of other components, while the activities of these components in turn help maintain the original agents. In this way, recurrent utilization creates self-stabilizing loops that resemble a minimal form of organizational closure. Functional relevance becomes a distributed property of the network rather than a property of any individual component.

\NI From this perspective, SBS may be viewed as a complementary principle connecting autocatalytic organization, organizational closure, and adaptive self-organization. Whereas autocatalytic and autopoietic theories emphasize the emergence of mutually sustaining structures, SBS emphasizes the local mechanism by which such structures are stabilized: components survive because they continue to serve functions for other components within the network.

\subsection{Core-Periphery Organization and Adaptive Exploration}

\NI One of the most striking findings of the present study is the spontaneous emergence of a stable core-periphery structure. A relatively small subset of agents becomes embedded in dense networks of mutual utilization and consequently receives sufficient reinforcement to suppress further adaptation. At the same time, a larger peripheral population remains weakly connected, receives little reinforcement, and therefore continues to explore alternative transformation rules. This separation emerges without explicit design or topological constraints, arising solely from local SBS dynamics.

\NI Such a division naturally resembles the well-known exploration--exploitation trade-off that appears throughout biological, cognitive, and technological systems. The core exploits previously discovered functional relations and thereby maintains stable system activity. The periphery, by contrast, remains adaptive and continues to search for alternative configurations. As a result, stability and innovation are not mutually exclusive but become spatially distributed across different regions of the same network.

\NI This interpretation is supported by the observation that adaptation events occur almost exclusively in the periphery, whereas the core remains remarkably stable over long periods of time. The core therefore acts as a reservoir of established functionality, while the periphery serves as a source of novelty. Such a division of labor may be particularly advantageous in adaptive systems because it allows ongoing exploration without disrupting already functional structures.

\NI The resulting organization also provides a potential explanation for the coexistence of robustness and evolvability. Robust systems must preserve functional organization despite internal fluctuations and external perturbations, whereas evolvable systems must remain capable of generating novelty and adapting to changing conditions. These requirements are often viewed as conflicting. In the present model, however, they emerge naturally from the same local SBS rule. Stability is concentrated within the mutually reinforcing core, while adaptive flexibility remains concentrated within the periphery.

\NI Interestingly, the formation of the core does not require external evaluation. In our simulations, stable internally circulating interaction networks emerge even before evaluators become active. This suggests that adaptive systems may develop substantial internal organization independently of immediate external demands. Such internally maintained dynamics can then function as a preadaptive search process, generating a repertoire of states and transformations from which future adaptations may be recruited.

\NI More generally, the emergence of core-periphery organization suggests that SBS may provide a simple local mechanism through which complex adaptive systems simultaneously achieve stability, innovation, and long-term adaptability. Rather than being imposed externally, the distinction between exploitation and exploration emerges spontaneously from the utilization dynamics of the system itself.

\subsection{SBS as a Substrate-Independent Principle}

\NI One of the most remarkable aspects of the SBS principle is that it makes no assumptions about the physical nature of the participating components. The mechanism does not depend on whether the interacting entities are molecules, genes, cells, organisms, neurons, artificial agents, social actors, or technological modules. Instead, SBS operates entirely at the level of functional interactions and organizational relations \cite{maturana1980autopoiesis,varela1979principles,moreno2015biological}. The only requirement is that components generate outputs which may subsequently be utilized by other components, and that utilization influences future persistence or adaptation.

\NI This distinguishes SBS from many domain-specific theories that are tied to particular physical substrates or mechanisms. In the present framework, the identity of the components is largely irrelevant. What matters is the existence of a network of interactions in which some outputs become functionally relevant for the continued activity of other components. As a consequence, the same underlying principle may potentially operate across a wide range of biological, cognitive, social, and technological systems. Similar perspectives have been proposed in theories of biological autonomy and organizational closure, which emphasize the primacy of relational organization over material implementation \cite{moreno2015biological,montevil2015biological}.

\NI From this perspective, SBS can be viewed as a relational rather than a material principle. Functional relevance is not an intrinsic property of a component, but emerges through its position within a network of ongoing interactions. Components survive not because of what they are, but because of what they do for other components. Persistence therefore becomes linked to participation in larger patterns of collective organization. In this sense, SBS complements existing theories of self-organization by proposing a simple local mechanism through which functionally relevant structures can become selectively stabilized \cite{camazine2020self, holland1992complex}.

\NI This interpretation resonates with several recent developments in biology and cognitive science. Increasingly, researchers have emphasized that many properties traditionally associated with cognition, agency, or problem solving may emerge from networks of interacting components rather than from particular physical substrates. Examples range from cellular information processing and collective behavior in multicellular systems to distributed problem solving in biological tissues and artificial agent collectives \cite{levin2019computational, levin2021life}. In such cases, the relevant organizational principles often appear to transcend the specific material realization of the system.

\NI Particularly relevant in this context is the growing recognition that adaptive behavior can emerge across multiple levels of biological organization. Levin and colleagues have argued that goal-directedness, information processing, and problem-solving capabilities may be understood as properties of collective systems spanning scales from cells to organisms and beyond \cite{levin2019computational, levin2021life}. Within such a framework, the boundary between cognitive and non-cognitive systems becomes less dependent on substrate and more dependent on organization and dynamics. SBS is compatible with this perspective because it focuses exclusively on patterns of functional interaction and adaptive stabilization rather than on any specific implementation mechanism.

\NI We do not claim that SBS is sufficient to explain the full complexity of biological organization, cognition, or agency. However, the present results suggest that utilization-dependent persistence may represent a simple and broadly applicable mechanism through which functionally coherent structures emerge. If so, SBS could provide a common conceptual language for describing self-organization across traditionally separated domains, ranging from chemical and biological systems to artificial life, adaptive technologies, and collective intelligence \cite{bedau2003artificial}.

\NI More generally, the substrate-independence of SBS raises the possibility that similar organizational principles may underlie a wide variety of adaptive systems. From this viewpoint, the emergence of functional structure becomes less a question of the material from which a system is built and more a question of how interactions are organized, stabilized, and maintained over time \cite{maturana1980autopoiesis, moreno2015biological, levin2021life}. Future work will be required to determine the extent to which SBS represents a domain-specific mechanism of adaptive networks or a more universal principle of self-organization.

\subsection{Economic Systems as a Special Case}

\NI The present work was originally inspired by economic systems, where decentralized agents collectively generate highly organized patterns of production and exchange without centralized control. Classical economic thought has long emphasized that complex societal organization can emerge from local interactions among individuals pursuing their own objectives. Adam Smith famously described this phenomenon through the metaphor of the ``invisible hand'', arguing that coordinated economic activity can arise without central planning \cite{smith1776wealth}. Later, Hayek emphasized that such organization depends on distributed knowledge that cannot be fully collected or processed by any central authority \cite{hayek1945use}.

\NI However, one of the main conclusions of the present study is that the SBS principle is not fundamentally economic. Rather, economic systems can be viewed as one particular realization of a more general mechanism. In our model, prices, utility functions, strategic planning, rational choice, and explicit optimization are entirely absent. The only feedback signal is whether the outputs generated by an agent continue to be utilized by other components of the system. Thus, the model strips away many domain-specific aspects of economics and retains only a minimal mechanism linking persistence to functional relevance.

\NI From this perspective, economic systems become one example among many systems in which SBS-like dynamics may operate. Markets provide a particularly intuitive illustration because the corresponding feedback signals are easily observable: products that are repeatedly purchased tend to persist, whereas products that fail to find users disappear or are modified. Yet the same underlying logic may also apply to biological, cognitive, technological, and social systems. What differs across domains is not the principle itself, but the physical realization of the feedback signals and the mechanisms through which adaptation occurs.

\NI This reinterpretation also helps explain why the model exhibits behaviors that extend beyond traditional economic questions. The spontaneous emergence of core-periphery structures, adaptive exploration, novelty generation, and self-sustaining interaction networks are not uniquely economic phenomena. Similar patterns have been reported in biological evolution, autocatalytic systems, neural networks, technological innovation, and other complex adaptive systems \cite{eigen1979hypercycle,kauffman1993origins,hordijk2018autocatalytic,borgatti2000models,csermely2013structure}. In this sense, the economic interpretation of the model may be viewed as a useful demonstration environment rather than its primary theoretical significance.

\NI Modern complexity economics already moves in a similar direction by treating economies as evolving adaptive systems composed of heterogeneous interacting agents rather than as equilibrium-based optimization problems \cite{arthur2021foundations}. Agent-based computational economics has further demonstrated how large-scale organization can emerge from local interactions \cite{tesfatsion2002agent, epstein1996growing}. The present work can be understood as an attempt to abstract one possible organizing principle from this broader tradition and place it within a more general framework of self-organization.

\NI Viewed in this way, economics occupies an important but ultimately secondary role within the SBS framework. Economic systems provide a particularly accessible and intuitive example of utilization-dependent persistence, but they are not the reason why the principle exists. Rather, they represent one domain in which a more general mechanism of self-organization becomes especially visible. The broader hypothesis proposed here is therefore not that economies operate according to SBS, but that economies are one manifestation of a principle that may apply across many classes of complex adaptive systems.

\subsection{Cognition, Artificial Life and Adaptive Systems}

\NI Beyond its relevance for self-organization and adaptive networks, the SBS principle may also have implications for the study of cognition and artificial life. Increasingly, cognition is viewed not as a property tied exclusively to nervous systems, but as an emergent feature of adaptive systems capable of maintaining functional organization, processing information, and flexibly responding to changing conditions \cite{maturana1980autopoiesis,moreno2015biological,levin2019computational}. From this perspective, the emergence and stabilization of functional roles becomes a central problem that spans biological, cognitive, and artificial systems alike.

\NI SBS offers a simple local mechanism through which such functional roles may emerge. Components that repeatedly contribute to ongoing system dynamics become stabilized, whereas components that fail to establish useful interactions remain subject to adaptation. Importantly, these roles are not specified externally. Rather, they emerge from the collective dynamics of the system itself. In this sense, SBS provides a mechanism by which functionally differentiated structures can arise without centralized control or explicit global optimization.

\NI The simulations presented here illustrate this process. Agents gradually discover transformations that become integrated into larger networks of mutually reinforcing interactions. Over time, some components acquire stable functional roles, while others continue to explore alternative configurations. Such dynamics resemble a distributed process of self-organization in which functionality is not imposed from outside, but emerges through recurrent patterns of utilization and reinforcement.

\NI The Missing Dimension Hurdle experiment highlights a particularly interesting aspect of this process. In this scenario, the system cannot satisfy the externally imposed target conditions through direct combinations of the available raw materials. Instead, successful states emerge only after repeated cycles of transformation and adaptation. The resulting dynamics resemble an open-ended search process in which the system progressively explores previously inaccessible regions of state space. Similar ideas appear in studies of innovation, exaptation, and the adjacent possible, where novel functions emerge through the recombination and reuse of existing components rather than through direct optimization toward a predefined goal \cite{gould1982exaptation,kauffman2000investigations,arthur2009nature}.

\NI From a cognitive perspective, such processes may be interpreted as the emergence of adaptive internal structure. Components that repeatedly participate in successful interaction patterns become increasingly stable and can subsequently serve as building blocks for higher-order organization. Although the present model does not explicitly learn representations, it demonstrates how recurrent utilization can give rise to increasingly structured repertoires of states and transformations. In this sense, SBS may provide a simple mechanism through which adaptive representations emerge from self-organized dynamics rather than explicit design.

\NI These observations also connect SBS to long-standing questions in artificial life. Artificial-life research seeks minimal mechanisms capable of generating adaptive behavior, novelty, and increasing organizational complexity \cite{bedau2003artificial,kauffman1993origins,hordijk2018autocatalytic}. The present results suggest that utilization-dependent persistence may represent one such mechanism. By continuously balancing stabilization and exploration, SBS enables systems to preserve established functionality while simultaneously searching for new functional configurations.

\NI We do not claim that SBS is sufficient to explain cognition, intelligence, or agency. However, the present findings suggest that it may contribute to a broader understanding of how adaptive systems acquire functional organization through local interactions. In this sense, SBS may provide a useful conceptual bridge between self-organization, artificial life, adaptive behavior, and the emergence of increasingly complex forms of information processing.

\subsection{Relation to Neural Networks and NeuroAI}

\NI The SBS principle is not restricted to the agent-based system studied here and may also be relevant for adaptive information-processing systems. In particular, several aspects of the present model suggest interesting connections to recurrent neural networks, reservoir computing, and self-organizing neural architectures. From this perspective, neural systems can be viewed as one possible realization of the more general SBS principle rather than as its primary motivation.

\NI A particularly interesting analogy exists to reservoir computing and recurrent neural networks. In the present model, raw materials are continuously supplied to a shared interaction space, a network of agents transforms these materials into intermediate states, and evaluators selectively extract states that satisfy specific target conditions. In a reservoir computer, external input signals are injected into a recurrent network, the reservoir transforms these inputs into a rich set of internal activity states, and a readout layer extracts task-relevant information from this dynamical repertoire \cite{jaeger2001echo,maass2002real,lukosevicius2009reservoir}. In both cases, a relatively unspecific internal system generates a diversity of intermediate states, while functional relevance is ultimately determined downstream. This connection is particularly natural in view of our previous work on recurrent neural networks, where we studied how network structure, recurrence, weight statistics, dynamical regimes, information flux, nonlinear dynamics, and organizational regularities shape the computational possibilities of recurrent systems \cite{krauss2019analysis,krauss2019recurrence,krauss2019weight,metzner2021dynamical,metzner2022dynamics,metzner2024quantifying,metzner2024recurrence,metzner2025nonlinear,metzner2025organizational}.

\NI The analogy can be formulated as a relation between state transformations and information processing. The present adaptive network transforms state vectors through chains of interactions, whereas a recurrent neural network transforms input signals through activity trajectories. Raw materials correspond loosely to input signals, the self-organized interaction network corresponds to the reservoir or recurrent core, and evaluators correspond to a readout mechanism that selectively utilizes what the internal system provides. In both cases, the internal network does not need to know the final task in detail. It can generate a broad repertoire of possible transformations, some of which may later become useful for downstream evaluation. This is close to the usual reservoir-computing intuition, but the present model makes the intermediate transformations more explicit: they appear as identifiable state-transfer events rather than as distributed state changes in a high-dimensional activity space.

\NI At the same time, the analogy has clear limits. A classical reservoir computer is usually initialized randomly and then kept fixed, whereas the agents in the present model adapt when they do not receive reinforcement. Moreover, states in the interaction space are discrete and persistent objects, while reservoir states are transient patterns of activity. The readout of a reservoir is often trained by a supervised target signal, whereas evaluators in our model merely select acceptable states and thereby provide a sparse local reinforcement signal. Nevertheless, the comparison may be useful precisely because of these differences. It suggests that one could study self-organizing reservoirs in which internal units, synapses, or modules are stabilized when their activity patterns are actually utilized by downstream parts of the network. In such a system, recurrence would not merely provide a fixed dynamical substrate, but would become an adaptive interaction network for information-processing functions.

\NI This perspective is also related to self-organizing recurrent networks such as SORN models \cite{lazar2009sorn}. In SORN-like systems, local plasticity rules such as spike-timing-dependent plasticity, intrinsic plasticity, and synaptic normalization can generate useful recurrent structure without explicit global error propagation. Such mechanisms are attractive because they show how recurrent networks can self-organize from local rules. However, their relation to a final computational task is often indirect. Standard backpropagation provides a complementary strategy: it assigns reinforcement by propagating error gradients from the output layer through the entire computational graph \cite{rumelhart1986learning,lecun2015deep}. This is highly effective, but it requires a global objective function and a structured backward reinforcement mechanism.

\NI The SBS principle suggests a possible intermediate route. Instead of asking whether a unit or connection directly reduces a global error, one may ask whether its output is utilized by other units that are themselves functionally relevant. A neural version of SBS could therefore assign a local usefulness variable to neurons, synapses, or modules. This variable would increase when the activity generated by the corresponding component contributes to downstream activity, memory, prediction, or readout performance. Components with high usefulness would become more stable, while components that remain unused, or that only participate in internally circulating activity without downstream relevance, would remain more plastic. A recursive version of the rule could propagate usefulness backward through actually utilized pathways, in analogy to the recursive reinforcement extensions discussed in the outlook below.

\NI Such an SBS-based learning rule would combine aspects of local self-organization and task-directed reinforcement. Like SORN-like rules, it would rely on local observables and could operate continuously during spontaneous or input-driven activity. Like backpropagation, it would introduce a directionality from final usefulness back into earlier processing stages. However, the signal would not have to be a precise gradient. It could be sparse, event-based, and functional: a component is stabilized because its output has acquired value for other components. This may be particularly relevant for modular recurrent systems, where the goal is not only to optimize a fixed input-output mapping, but also to allow useful intermediate functions to emerge, stabilize, and become reusable building blocks of later computation.

\NI In this sense, SBS may provide a conceptual bridge between theories of self-organization in complex adaptive systems and future learning mechanisms for adaptive neural architectures. Rather than relying exclusively on global optimization, such systems could exploit local signals of functional relevance to stabilize useful internal structure while maintaining exploratory capacity.

\subsection{Outlook}

\NI In the present work, we have used selected parameter combinations to test the basic SBS hypothesis and to demonstrate that utilization-dependent persistence can generate self-organized interaction networks. However, the full parameter space of the model remains largely unexplored. A natural next step is therefore a systematic parameter study. Particularly interesting control parameters include the acceptance threshold, which determines how easily states can enter new transformation steps; the out-of-subspace contribution, which controls the generation of genuinely novel state properties; the adaptation time scale and mutation amplitude, which regulate the balance between stability and exploration; and the state-aging probability, which controls how long unused states remain available for later recombination. It would also be important to vary the number of agents, the dimensionality of state space, the number and diversity of raw materials, and the number of evaluators. Such studies could reveal whether the emergence of stable cores, adaptive peripheries, and self-sustaining interaction networks represents a robust dynamical pattern of SBS systems or is restricted to particular parameter regimes.

\NI Another important extension concerns the reinforcement signal itself. In the current model, an agent receives reinforcement whenever one of its outputs is utilized, either by another agent or by an evaluator. This simple rule is sufficient to generate internally organized interaction networks, but it can also stabilize cores that mainly reinforce themselves and are only weakly coupled to externally relevant outcomes. A natural extension would therefore distinguish between ordinary utilization reinforcement and relevance-weighted reinforcement. Each agent could carry a slowly changing estimate of its downstream relevance, which is high when the agent contributes directly to successful evaluations and which can be inherited, in discounted form, by upstream agents through actual state-transfer events. In this way, functional relevance could propagate recursively backward through the network without requiring any agent to possess knowledge of the global interaction structure. Such a recursive reinforcement mechanism would preserve the exploratory value of internally circulating networks while gradually biasing long-term stability toward pathways that contribute to externally relevant outcomes.

\NI More generally, future work could investigate whether SBS remains effective in substantially richer environments. The present model employs fixed agent populations, simple transformation rules, and externally imposed target conditions. Open-ended systems may instead allow the continuous generation of new agents, new transformations, and new forms of interaction. Under such conditions, SBS might contribute not only to self-organization but also to the emergence of increasingly complex adaptive structures. This possibility connects the present work to long-standing questions in artificial-life research concerning open-ended evolution, novelty generation, and the spontaneous emergence of higher levels of organization \cite{bedau2003artificial,kauffman1993origins,hordijk2018autocatalytic}.

\NI Another promising direction concerns the relation between SBS and cognition. The present model does not contain explicit representations, memory systems, or goal-directed behavior. Nevertheless, the observed dynamics suggest that stable functional roles can emerge through purely local interactions. It would therefore be interesting to investigate whether more sophisticated SBS systems can give rise to adaptive internal representations, predictive structures, or other forms of self-organized information processing. Such work could help clarify the extent to which cognitive phenomena depend on specific substrates or whether they can emerge more generally from patterns of functional interaction and adaptive stabilization \cite{levin2019computational,moreno2015biological}.

\NI Finally, the SBS principle may prove useful as a conceptual framework for comparing adaptive systems across traditionally separate disciplines. The same underlying mechanism could potentially be explored in chemical reaction networks, biological systems, neural architectures, artificial agents, technological infrastructures, and social organizations. If the central idea survives these tests, SBS may represent not merely a property of a particular model, but a more general principle linking self-organization, functional relevance, adaptation, and the emergence of complex structure across multiple domains.

\section{Additional Information}

\subsection{Author contributions}

CM supervised the study, implemented simulations, evaluated reults, and wrote the paper. AG implemented simulations and evaluated results. AS discussed the results and acquired funding. AM and TK discussed the results and provided resources. PK conceived the study, discussed the results, acquired funding and wrote the paper.

\subsection{Funding}
This work was funded by the Deutsche Forschungsgemeinschaft (DFG, German Research Foundation): grants KR\,5148/3-1 (project number 510395418), KR\,5148/5-1 (project number 542747151), KR\,5148/10-1 (project number 563909707) and GRK\,2839 (project number 468527017) to PK, and grants SCHI\,1482/3-1 (project number 451810794) and SCHI\,1482/6-1 (project number 563909707) to AS.

\subsection{Acknowledgements}
For the publication fee we acknowledge financial support by Heidelberg University.

\subsection{Competing interests statement}
The authors declare no competing interests.

\subsection{Data availability statement} 
The complete data and analysis programs will be made available upon reasonable request.

\subsection{Third party rights}
All material used in the paper are the intellectual property of the authors.


\bibliographystyle{unsrt}
\bibliography{references}


\newpage

\newpage

\newpage

\newpage

\newpage

\newpage

\end{document}